# Symmetry of the flows of Newtonian and non-Newtonian fluids in the diverging and converging plane channels

Alexey I. Fedyushkin, Artur A. Puntus, and Evgeny V. Volkov

View Online    Export Citation

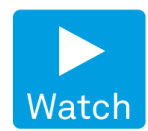

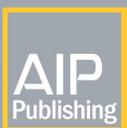

# Symmetry of the Flows of Newtonian and Non-Newtonian Fluids in the Diverging and Converging Plane Channels

Alexey I. Fedyushkin[1,a)], Artur A. Puntus[2] and Evgeny V. Volkov[1,2,b)]

[1]*Ishlinsky Institute for Problems in Mechanics of Russian Academy of Sciences, Moscow 119526, Russia*
[2]*Moscow Aaviation Institute, Volokolamsk highway, d.4, Moscow 125993, Russia,*

[a)]Corresponding author: fai@ipmnet.ru
[b)]evvolkov94@mail.ru

**Abstract.** The results studying various laminar flow regimes in diverging and converging plain channels (diffuser and confusor) with a small opening angle of channels (diverging and converging angles) are presented. The results are obtained for a viscous incompressible fluid by numerical simulation based on solving the Navier-Stokes equations. The paper presents the results concerning the change in the nature of flows from stationary - symmetric to stationary - asymmetric and to non-stationary in the diffuser and confusor in dependence on the Reynolds number. The ranges of existence of these flow regimes in plane diffusers and confusors depending on the Reynolds number for Newtonian, pseudo plastic and dilatants fluids with the Ostwald-de Waele power law for viscosity are numerically found. The transitions of flow regimes in the diffuser from symmetric steady state to the asymetric one and to the asymetric unsteady mode in dependence on the Reynolds number are shown. The values of Reynolds number that determine the existence ranges of these flow modes in the cases of Newtonian and non-Newtonian fluids are given.

## INTRODUCTION

In this paper we consider the problems of laminar flows of Newtonian and non-Newtonian viscous incompressible fluids in a diverging and converging plane channels (diffuser and confusor) with small an opening angles (angles of the diverging and converging channels). This problem has not only of fundamental scientific importance for hydro mechanics, but also it has an applied interest in numerous industries (aviation, space, heat power, oil and gas production, medicine, hydrology, etc.). The problem under consideration similar to the classical Jeffrey-Hamel problem (JH) in the case of a flow of a viscous incompressible fluid in a diffuser and a confusor at various Reynolds numbers [1,2]. The difference between our statement of the problem and the Jeffrey-Hamel one consists in the fact that we consider only a finite geometry of diverging or converging channels and our statement lies beyond the restriction of an existence of asymmetric flows, and it is studied numerically. During the flow of incompressible fluid in the diffuser and confusor, due to the expansion or compression of the flow area, the kinetic energy of the flow is converted into the potential energy of the pressure drop, which distinguishes one from the flow in the channel with parallel walls. In the case of diffuser – the flow towards the pressure gradient, and in the case of confusor – the flow is directed in in the same direction with the pressure gradient. These are two different situations, which radically affects the occurrence of bifurcations and asymmetry of the flow [3-5]. The JH problem is simple in formulation, but it is saturated with a lot of features, physical phenomena, has many applications and it is therefore urgent up to present day. There are many works devoted to studying incompressible fluid flows in diverging and converging plane channels (diffuser and confusor), for example see [1-22]. The details and features of the JH problem are described and discussed in manuscripts [3,4]. The paper [5] provides a review of more than 150 papers related to solving the JH problem and a generalization on the basis of group analysis of differential equations. Also in this paper [5], the asymptotic character of JH solutions is studied and general solutions are given. Early investigations JH flow were done in the paper [6] and the solution of the stationary two-dimensional JH problem expressed in terms of the elliptic Jacobi function was obtained. In the papers [7,8] the generalizations of the solution to the JH problem was the presence of asymmetrical stationary flow patterns was indicated, and one, two, and three mode bifurcation solutions were presented. These studies point to a presence of stationary asymmetric and multi-mode solutions for certain ranges of Reynolds numbers and opening angles of a diffuser.

The flows of jets into the open space and flows to suddenly expanding channels can be considered as a limiting case of flow in diffusers with opening angles of 360 or 180 degrees, respectively. The study of flow symmetry breaking and bifurcations in the JH problem with including the flow in the channel and the limiting case of a diffuser with an opening angle of 180 degrees, was carried out in [9]. The results of experimental and theoretical studies of the flow in a symmetric suddenly expanding channel are presented in [10-14]. The paper [10] presents flow patterns and velocity profiles in a channel with symmetric expansion measured by the laser - Doppler method. In paper [10] it is shown experimentally that at small Reynolds numbers in the symmetric channel with step expansion the flow can be of stationary and asymmetric character. In [11] it was shown that at low Reynolds numbers the energy fluctuations in the channel can exceed the energy fluctuations caused by turbulence. The papers [11-17] present the study of bifurcation occurrence in the JH fluid flows are the critical values of the Reynolds numbers of occurrence of various types of bifurcations and neutral curves (Reynolds number vs aspect ratio of the shoulder) for a stationary, periodic, symmetric, and asymmetric flows in channels with sudden expansion. The appearance of stationary and non-stationary waves in the fluid flow in the diffuser within the range of Reynolds numbers from 5 to 5000 at various opening angles of the diffuser was studied in the paper [15]. The dependences and asymptotics of the existence of waves depending on the opening angle of the diffuser and Reynolds number are given. In the paper [17] the results of 2D and 3D (under the condition of symmetry) numerical of a viscous incompressible fluid flow at Reynolds numbers within the range from 60 to 360 and opening angle from 10° to 180° in a plane diffuser with the presence of the input plot shows the effect of three-dimensionality and oscillations of the velocity at the inlet of the diffuser to the presence of non-steady asymmetric flow regimes. Computational results [18] indicate that temporally stable, isolated, steady solutions may exist for finite-domain analogues of the steady waves presented in [15] for an infinite domain. In [18] it was demonstrated that there is non-uniqueness of stable solutions in a certain parameter regime.

The extension of the traditional JH problem to flows in flat channels with stretchable convergent / divergent walls was done in [19,20]. In these works, the influence of opening angle and the rate of stretching walls on the main flow was shown. The problem of numerical simulation of unsteady subsonic viscous gas flows in a channel with a sudden symmetric expansion of the cross section (diffuser) was considered in the paper [21]. In a wide range of characteristic parameters, nonlinear processes of instability development of the flow under consideration have been numerically studied with allowance for acoustic-vortex interactions. The effects of a sound self-excitation of a jet flowing into a wide part of the channel were discovered. The influence of the input profiles of the average velocity on the flow evolution was evaluated.

It is known that the nature of a viscous fluid flow in a plane two-dimensional diffuser / confusor is determined by the geometry (the opening angle of diverging or converging channels β) and the Reynolds number. The opening angle of the diffuser begins to have a significant impact on the flow characteristics when that exceeds ten degrees [7,15,18], therefore, in this paper, the simulation was performed for diffusers and confusors with a small opening angle of diverging and converging of channel. The paper [22] presents the results based on numerical solving the JH problem for diffuser with a small opening angle. Changing modes of flow in a diffuser depending on the Reynolds number from a symmetric stationary to asymmetrical stationary and to non-stationary asymmetric were demonstrated. The values of ranges of Reynolds numbers are specified for the existence of these modes. Our paper is a continuation of the paper [22] and presents the results concerning the change in the character of flows from stationary - symmetric to stationary - asymmetric and to non-stationary in a diffuser and confusor in dependence on the Reynolds number. In [22] numerical method of control volumes was used, which was described in [23]. The existence ranges of these flow regimes in plane diffusers and confusors are numerically determined in dependence on Reynolds number (flow velocity) for Newtonian, pseudoplastic and dilatant fluids with the Ostwald - de Waele power law for viscosity. It was of interest to see how the character of flows changes for rheologicaly complex fluids, for example, a fluid that satisfies this power law [24,25]. In present paper our results of comparing numerical modeling of laminar flows of a viscous fluid in a plane diffuser and confusor under symmetric and asymmetric boundary conditions at the entrance are presented. An analysis of the results obtained enable us to conclude that there exists a significant difference in the regimes of symmetrical fluid flows in the diffuser and confusor.

# THE PROBLEM STATEMENT

The laminar flow of a viscous incompressible fluid driven through a channel bounded by two flat walls inclined towards each other at a small angle β is considered. In this paper we consider plane long but bounded channels with two arcs ("input" and "output") with the same center (Fig. 1, 2) unlike to the classical problem of Jeffrey – Hamel symmetry flow in a plane dimensionless diffuser (confusor). The purpose of our numerical simulations is a determination of the existence ranges of symmetric and asymmetric stationary flows and the transition to the non-stationary mode (for pseudoplastic, Newtonian and dilatant fluids in the range of Reynolds numbers Re <500) and compare the obtained simulation results.

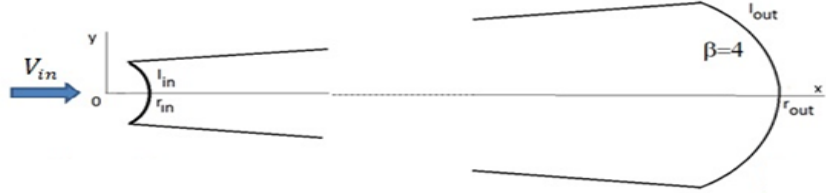

**FIGURE 1.** Scheme of the computational domain for a plane diffuser ($\beta = 4°, L = 0.495$ meter)

Geometric model of the diffuser is as follows: opening angle is $\beta = 4°$, the input boundary has the form of an arc $l_{in}$ ($r_{in} = 0.005$ meters) where $r$ is calculated by formula (1)

$$r^2 = x^2 + y^2 \tag{1}$$

The output boundary has the form of an arc $l_{out}$ ($r_{out} = 0.5$ meters) Fig. 1. The length of the diffuser $L$ is equal to 0.495 meters, and is calculated by formula (1.2)

$$L = |r_{out} - r_{in}| \tag{2}$$

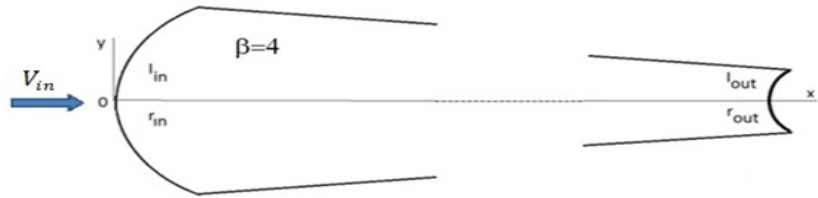

**FIGURE 2.** Scheme of the computational domain for a plane confusor ($\beta = 4°, L = 0.495$ meter)

Geometric model of confusor: angle of the converging β = 4°, the input boundary has the shape of an arc $l_{in}$ ($r_{in} = -0.5$ meters), where $r$ is also calculated by formula (1), and the output boundary has the shape of an arc $l_{out}$ ($r_{out} = -0.005$ meters) Fig. 2. The length of the confusor $L$ is 0.495 meters and is calculated by formula (2).

The simulation of the problem is carried out on the basis of the numerical solving the two-dimensional Navier – Stokes equations (3) for an incompressible viscous fluid

$$\frac{\partial V}{\partial t} + (V\nabla)V = -\frac{\nabla P}{\rho} + \frac{1}{\rho}\nabla\sigma(V)$$
$$div\, V = 0 \tag{3}$$

where $V = V(v_x, v_y)$ is the velocity vector; $P$ – the pressure; $\rho$ –the density; $\sigma(V)$– the viscous stress tensor.

As the boundary conditions we take: at the arc of inlet to the diffuser / confusor, velocity is a constant which value is calculated from the positive mass flow rate $Q_{in}$ (Reynolds number $Re_{in}$), at the arc of output the pressure is assumed $P = 0$. At the upper and lower walls for velocity the no slip conditions are $V = 0$ is assumed. The initial conditions are $t = t_0 = 0$, $V(t_0) = 0$, $P = 0$.

Numerical calculations were carried out using the method of control volumes [23] with schemes of increased accuracy in space and first order in time. Test calculations were carried out on sequence of the meshes with decreasing grid step. The number of grid nodes was chosen, on which the results did not differ as it increases. The results for stationary and quasi-stationary flow regimes are obtained on grids with a margin (Reynolds grid numbers were not more than one as well as the values of Courant numbers). The grids were rectangular and orthogonal on solid walls. We used grids with decreasing grid spacing at the input and output of the region, as well as near solid boundaries (there were at least ten grid nodes in the boundary layers). On solid walls (in boundary layers) the mesh was orthogonal on a distance of several tens of mesh steps. In addition, to determine the sufficiency of the mesh size, velocity profiles along solid walls along lines parallel and at a small distance from the solid walls (0.1 of the size of the inlet part of the diffuser and parallel to the solid walls) were compared for different grids. For a given input velocity, calculations were performed from zero initial velocities in the computational domain until a steady-state (or quasi-steady-state) flow regime was established. The analyzing of the numerical solutions and the nature of the flow was carrying out at the steady-state time regime.

To describe the rheology of a wide class of non-Newtonian fluids [24,25] the Ostwald-de Waele power law for was used and the shear stress τ was given by (4)

$$\tau = k\gamma^n \qquad (4)$$

where: $k$ is the flow density coefficient (SI units Pa s$^n$): the greater is $k$, the less is the medium mobility, $\gamma$ is the shear rate or the velocity gradient normal to the plane of shear (SI unit s$^{-1}$), and $n$ is the flow behavior index (dimensionless). The parameters $k$ and $n$ are rheological parameters, and they are constant for a fluid in limited range of shear rates. The more strongly $n$ differ from unity, the more pronounced is the viscosity anomaly and the nonlinearity of the flow curve.

The simulation was carried out for Newtonian, pseudoplastic and dilatant fluids with the Ostwald-de Waele power law for viscosity [24]. In the one dimensional case the viscous stress for this law can be written as:

$$\tau = k\left(\frac{\partial u}{\partial y}\right)^n \qquad (5)$$

where $k > 0$–is the flow density coefficient (in SI the unit of measure is $Pa \cdot s^n$), $(\partial u/\partial y)$ - is the shear rate or the velocity gradient normal to the plane of shear (SI unit $s^{-1}$) (in SI measured in $s^{-1}$), n - is the flow behavior index (dimensionless). The value of viscosity defined as a function of the velocity gradient by formula (6):

$$\mu_{eff} = k\left(\frac{\partial u}{\partial y}\right)^{n-1} \qquad (6)$$

$\mu_{eff}$ is the effective viscosity measured in $Pa \cdot s$ in SI. In the two-dimensional case, the components of viscous stress tensor for the Ostwald-de Waele power law can be found, (see for example, [24,25]). The types of fluids with viscosities corresponding to the law of viscosity (6) are given in Table 1 [25].

**TABLE 1.** The types of fluids depending on the exponent n for viscosity law (9).

| n | Fluid type |
|---|---|
| n < 1 | Pseudoplastic |
| n = 1 | Newtonian |
| n > 1 | Dilatant |

The rheological parameters $k$ and n are determined from experiments and analysis of consistency curves. For real liquids the values of n and $k$ are are not constant for a great values of shear stress. However, the fluids obeying this law have wide practical applications since in practice one deals with a limited range of shear rates [24,25].

The numerical solving the two – dimensional Navier-Stokes equations was carried out by the method of control volumes [23], the schemes of the second and third order accuracy in space and the first order accuracy in time were used. Test calculations were carried out on the sequence of grids with decreasing grid spacing. From these tests, the calculated grids with such a number of grid nodes were selected, with the increase of which the results did not differ. Numerical calculations were obtained on grids with a margin of accuracy to prevent the emergence of numerical instability (the values of Reynolds grid numbers and the values of Courant numbers were not more than unit). Detailed non-uniform grids were used in the modeling and accuracy control was carried out at each time step. The grid cells near solid longitudinal walls were rectangular and orthogonal. Non-uniform grids were used with decreasing grid spacing near the entrance and exit to the diffuser or confusor, as well as near solid longitudinal boundaries (there were at least ten grid nodes in the boundary layers). Near the solid walls (in the boundary layers), the grid was orthogonal at a distance of several tens of grid steps. For a given velocity at the entrance to the diffuser or confusor, the calculations were performed from zero initial velocities in the computational domain to the establishment of a stationary (or quasi-stationary) flow regime. The analysis of numerical results and the nature of the flow were carried out on a steady-state stationary or quasi-stationary mode.

## RESULTS

In this section we present the numerical results of numerical simulation for a fluid that satisfies the Ostwald-de Waele power law, namely, pseudoplastic, dilatant, and Newtonian. The simulation results for the diffuser and for the confusor are shown in Table 2 and in Table 3 correspondingly. In Tables 2, 3 the flow's types are denoted by the following abbreviations: **ssf**-stationary symmetric flow, **saf** - stationary asymmetric flow, **nsaf** - non-stationary asymmetric flow. Fluid flows in the diffuser and the confusor differ significantly from each other in terms of their symmetry breaking. This can be seen from the comparison of the results from Table 2 and Table 3.

**TABLE 2.** The results of numerical simulation for a diffuser.

| | | | | |
|---|---|---|---|---|
| **Pseudoplastic (n=0.5)** | | | | |
| $Q_{input}$[kg/s] ($Re_{in}$) | 0.05 (50) | 0.1 (100) | 0.3 (300) | 0.5 (500) |
| Flow type | **ssf** | **ssf** | **saf** | **nsaf** |
| **Newtonian (n=1)** | | | | |
| $Q_{input}$[kg/s] ($Re_{in}$) | 0.05 (50) | 0.1 (100) | 0.3 (300) | 0.5 (500) |
| Flow type | **ssf** | **ssf** | **saf** | **nsaf** |
| **Dilatant (n=2)** | | | | |
| $Q_{input}$[kg/s] ($Re_{in}$) | 0.05 (50) | 0.1 (100) | 0.3 (300) | 0.5 (500) |
| Flow type | **ssf** | **ssf** | **ssf** | **ssf** |

**TABLE 3.** The results of numerical simulation for confusor.

| | | | |
|---|---|---|---|
| **Pseudoplastic (n=0.5)** | | | |
| $Q_{input}$[kg/s] ($Re_{in}$) | 0.05 (50) | 0.1 (100) | 0.5 (500) |
| Flow type | **ssf** | **ssf** | **ssf** |
| **Newtonian (n=1)** | | | |
| $Q_{input}$[kg/s] ($Re_{in}$) | 0.05 (50) | 0.1 (100) | 0.5 (500) |
| Flow type | **ssf** | **ssf** | **ssf** |
| **Dilatant (n=2)** | | | |
| $Q_{input}$[kg/s] ($Re_{in}$) | 0.05 (50) | 0.1 (100) | 0.5 (500) |
| Flow type | **ssf** | **ssf** | **ssf** |

In Figs. 3-5 for the diffuser the dimensionless profiles of the velocity vector modulus: $V_{\_dimless} = \frac{\sqrt{v_x^2+v_y^2}}{V_{in}}$ vs dimensionless coordinate: $y_{\_dimless} = \frac{y}{r}\sin\left(\frac{\beta}{2}\right)$ in four vertical cross sections ($x$ = 0.1, 0.2, 0.3, 0.4 m) are shown.

The results of calculations of stationary asymmetric flows, obtained for the diffuser, confirm show that at a certain value of Re there is a violation of symmetry in the structure of the flow, in addition, vortex structures with return opposite reverse flows are formed near the walls, this is a well-known fact from numerous literature. Numerical calculations showed that as the Reynolds number exceeds the critical value $Re > 269$, the boundary layer breaks down and at a certain distance from the entrance to the diffuser, weak return flows are formed near the longitudinal walls of the diffuser, both for $n = 0.5$ and $n = 1$. In these areas, the velocity vector changes direction.

Numerical calculations showed that the flow in the diffuser ceases to be symmetric, remaining stationary in a narrow range of Reynolds numbers $300 > Re > 279$ for different types of fluid.

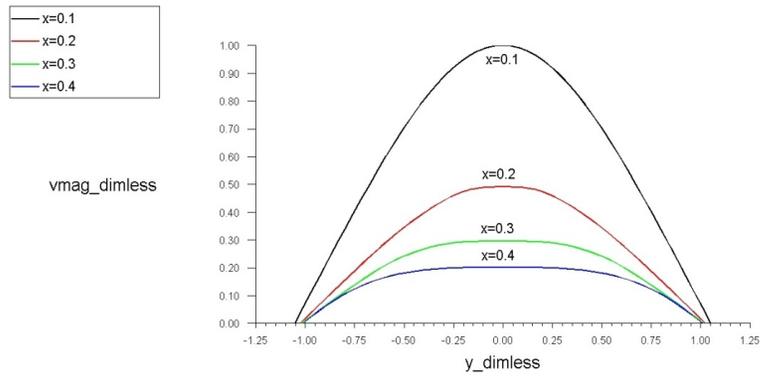

**FIGURE 3.** Dimensionless profiles of the velocity vector modulus for a pseudoplastic fluid for $Q = 0.1\ [kg/s]$, $n = 0.5$ (in cross sections: x = 0.1, 0.2, 0.3, 0.4 m)

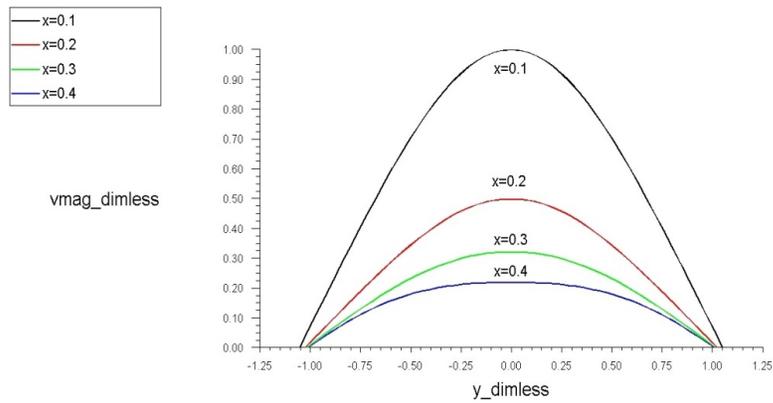

**FIGURE 4.** Dimensionless profiles of the velocity vector modulus for a Newtonian fluid for $Q = 0.1\ [\frac{kg}{s}]$, $n = 1$ (in cross sections: x = 0.1, 0.2, 0.3, 0.4 m)

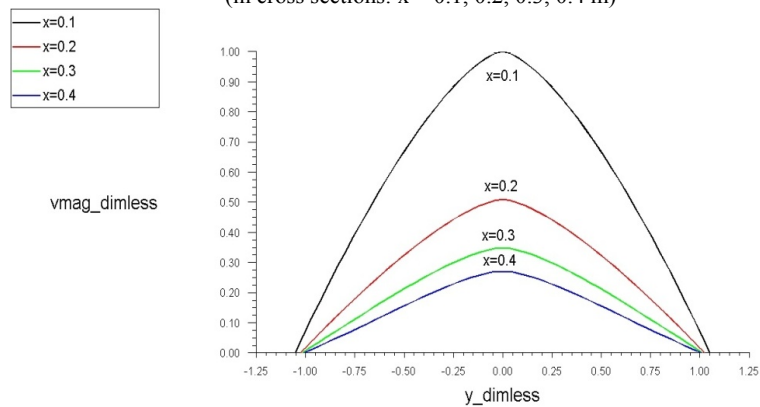

**FIGURE 5.** Dimensionless profiles of the velocity vector modulus for a dilatant fluid for $Q = 0.5\ [\frac{kg}{s}]$, $n = 2$ (in cross sections: x = 0.1, 0.2, 0.3, 0.4 m).

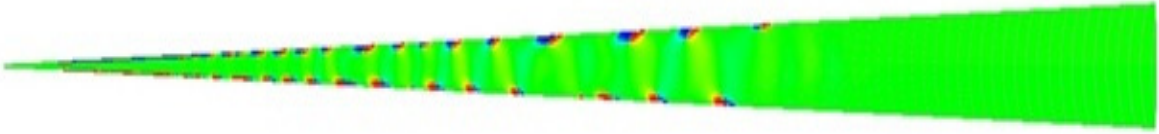

**FIGURE 6.** Isolines of angle of the velocity vector for a dilatant fluid in the diffuser
($Q = 0.3 [kg/s]$, $Re = 300$, $n = 2$)

In Fig. 6 for diffuser the color isolines of the angle of velocity vector ($\arctg(v_x/v_y)$) with a stationary symmetric type of flows a dilatant fluids (for $Re = 300$, $n = 2$) are shown.

As the Reynolds number increases, the intensity of the near-wall secondary vortex flows increases, the pattern of their arrangement in the longitudinal direction have formed a "chess" structure, and the main flow has the form of a jet alternating between the continuous longitudinal walls of the diffuser in time. As the Reynolds number increases the stationary flows in the diffuser lose their stability and pass into the oscillatory flow region, which characteristics vary along the length of the diffuser in time.

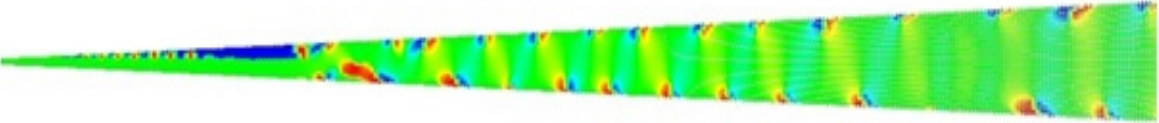

**FIGURE 7.** Isolines of the angle of velocity vector for pseudoplastic fluid ($Q = 0.5 [kg/s]$, $Re = 500$, $n = 0.5$)

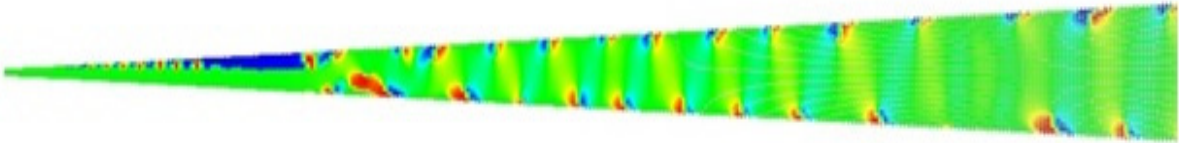

**FIGURE 8.** Isolines of the angle of velocity vector for Newtonian fluid ($Q = 0.5 [kg/s]$, $Re = 500$, $n = 1$)

In Figs. 7 and 8 for diffuser the color isolines of the angle of velocity vector ($\arctg(v_x/v_y)$) with a non-stationary asymmetric type of flows for pseudoplastic and Newtonian fluids (for $Re = 500$, $n = 0.5$ and $n = 1$) are shown. The results show the chess intermittent structure of the unsteady flow.

## CONCLUSIONS

For pseudoplastic and Newtonian fluids in a plane diffuser, the range of existence of stationary symmetric, stationary asymmetric and non-stationary asymmetric flow regimes is numerically shown. Numerical results showed that as the Reynolds number exceeds the critical value $Re > 269$, the boundary layer breaks down and at a certain distance from the entrance to the diffuser, weak return flows are formed near the longitudinal walls of the diffuser, both for $n < 1$ and $n = 1$.

Numerical calculations showed that the flow in the diffuser ceases to be symmetric, remaining stationary in a narrow range of Reynolds numbers $300 > Re > 270$ for different types of fluid. For a dilatant fluid in the considered range of Reynolds numbers, only stationary symmetric flows were identified.

In the considered range of Reynolds numbers (Re <500) in a plane confusor, the currents are stationary and symmetric, a stationary mode with an asymmetric flow structure is not observed.